\documentclass[aps,prx,twocolumn,bibliography,superscriptaddress]{revtex4-1}
\usepackage{amssymb}
\usepackage{graphicx}
\usepackage{amsmath}
\usepackage{amsfonts}
\usepackage[dvipsnames]{xcolor}
\usepackage{amsthm}
\newtheorem*{theorem*}{Theorem}
\begin{document}

%\title{Unveiling Non-Hermitian Topological Phases and Engineering Non-Hermitian Flat Band Physics through Real Space Decimation}

\title{Non-Hermitian Topology and Flat Bands via an Exact Real Space Decimation Scheme}

\author{Ayan Banerjee}
\thanks{These authors contributed equally}
%\email{ayanbanerjee@iisc.ac.in}
\author{Arka Bandyopadhyay}
\thanks{These authors contributed equally}
\affiliation{Solid State and Structural Chemistry Unit, Indian Institute of Science, Bangalore 560012, India}
 \author{Ronika Sarkar}
\thanks{These authors contributed equally}
\affiliation{Solid State and Structural Chemistry Unit, Indian Institute of Science, Bangalore 560012, India}
\affiliation{Department of Physics, Indian Institute of Science, Bangalore 560012, India}
\author{Awadhesh Narayan}
\email{awadhesh@iisc.ac.in}
\affiliation{Solid State and Structural Chemistry Unit, Indian Institute of Science, Bangalore 560012, India}
\date{\today}

%\date{\today}

\begin{abstract}
In recent years, non-Hermitian phases in classical and quantum systems have garnered significant attention. In particular, their intriguing band geometry offers a platform for exploring unique topological states and unconventional quantum dynamics. However, their topological characterization becomes particularly interesting and challenging in complex multiband systems. Here we propose a decimation framework, which leverages real space renormalization group to streamline the analysis of complex multiband non-Hermitian systems. Our systematic approach allows us to probe different phases and transitions, analyze bulk-boundary correspondence, formulate generalized Brillouin zones, investigate open boundary spectra, survey non-Bloch van Hove singularities, study disorder-induced effects, and explore tunable non-Hermitian flat band physics. Additionally, our framework allows proposing a hypothesis about quasi-one-dimensional bipartite non-Hermitian systems with flat bands, demonstrating their decoupling into Su-Schrieffer-Heeger chains and compact localized states across various models. Our work presents a powerful and comprehensive framework for understanding the intricate properties of non-Hermitian multiband systems, offering insights into the evolving landscape of non-Hermitian topological physics.

\end{abstract}
\maketitle

\section{Introduction} 

The exploration of non-Hermitian phases in open classical and quantum systems has gained significant traction in both theoretical and experimental realms~\cite{bergholtz2021exceptional,ashida2020non,el2018non,zhang2022review,banerjee2023non}. Non-Hermiticity enriches and offers unique topological phases considering the interplay between ramified symmetries and topology~\cite{kawabata2019symmetry,ding2022non}. The presence of exceptional degeneracies in non-Hermitian systems leads to intriguing spectral topology, facilitating non-Hermitian phase transitions and introducing the notion of ``point" and ``line" gap topologies in the complex plane~\cite{gong2018topological,zhang2020correspondence}. The spectral topology and its extreme sensitivity to boundary conditions give rise to interesting phenomena such as the non-Hermitian skin effect (NHSE)~\cite{yao2018edge,yao2018non}. This, in turn, has led to exciting applications such as enhanced lasing~\cite{feng2014single}, topological funnelling of light~\cite{weidemann2020topological}, unique uni-directional transport~\cite{zhao2019non,longhi2015robust}, among several others. Non-Hermitian flat bands, characterized by their peculiar properties and correlations, offer a rich platform for exploring intriguing phenomena such as unique topological states, non-reciprocal transport, and unconventional quantum dynamics, challenging conventional band structures and expanding our understanding of correlated physics~\cite{longhi2015robust,wang2021topological,leykam2017flat}. On the other hand, the spectral topology and concomitant topological characterization in terms of an appropriate topological invariant become more interesting as well as challenging when the system has enhanced degrees of freedom and multiple $(n\geq 3)$ number of energy bands~\cite{teo2020topological,guo2023exceptional}. Recent progress in both theoretical and
experimental activities that explore the physics of non-Hermitian multi-band systems are quickly altering the research landscape~\cite{zhang2023experimental,tang2022experimental,wang2021topological,wang2021generating}. Given the recent developments, there is a need for a general approach that can offer a more profound theoretical understanding of these complex multiband systems. 

In this work, we propose a complementary non-Hermitian framework based on a decimation scheme~\cite{ashraff1988exact,chakrabarti1995role,bandyopadhyay2020review}, which serves as a backbone of the scale transformation in the real space renormalization group (RSRG) framework~\cite{kadanoff1966scaling,jagannathan2004quantum,angelini2017real,koch2018mutual}.
More specifically, our formalism is \emph{exact} and utilizes the power of renormalization group theory to integrate out the chosen degrees of freedom, resulting in the down-folding of a complex Hamiltonian. In other words, our method efficiently maps the original system to a smaller, simplified yet self-contained system, which retains the information regarding the band theoretic and associated topological properties through its coarse-grained renormalized parameters.

The formalism proposed here is a powerful tool for streamlining the analysis of complex multiband non-Hermitian systems. Our systematic approach has yielded intriguing results, including 
(i) Probing different phases and phase transitions of generalized non-Hermitian multiband models.
(ii) Analyzing the bulk-boundary correspondence (BBC) using a transfer matrix approach.
(iii) Formulating a generalized Brillouin zone (GBZ) for a complex multiband system.
(iv) Investigating open boundary spectra and skin modes.
(v) Surveying non-Bloch van Hove singularities.
(vi) Studying the effect of impurity and disorder in transport properties. 
(vii) Exploring tunable non-Hermitian flat band physics using a general prescription.
Additionally, our approach also sheds light on the qualitative understanding of compact localized states (CLS) in non-Hermitian systems. In particular, we hypothesize that \emph{any quasi-one-dimensional (Q1D) bipartite non-Hermitian system exhibiting a flat band can be decoupled into a non-Hermitian Su-Schrieffer-Heeger (SSH) chain and periodically arranged isolated sites.} The non-Hermitian SSH chain enables the band topology, while the latter manifests the non-dispersive band. Our hypothesis is tested for different Q1D non-Hermitian lattice models. Overall, our formalism offers a comprehensive framework for studying and understanding the properties of complex multiband non-Hermitian systems.

\section{Formalism}

The tight-binding Hamiltonian for a noninteracting fermionic system can be expressed in the tight-binding representation as

\begin{equation}
\hat{H} = \sum_n | n \rangle \tilde{\epsilon}_n \langle n | + \sum_{n \neq m} | n \rangle \tilde{V}_{nm} \langle m |,
    \label{eq:tightbinding}
\end{equation}

where the complex variables $\tilde{\epsilon}_n$ represent the onsite energies, while $\tilde{V}_{nm}$ denote the transfer (hopping) energy between the orbitals $|n \rangle$ and $|m \rangle$. Furthermore, the equation of motion for the Green's function can be derived using the tight-binding Hamiltonian provided and is expressed as $\sum_{l} (E \delta_{nl} - H_{nl})G_{lm}(E) = \delta_{nm}$. In the case of non-Hermitian systems, the retarded electronic Green's function $G^R$ has the form $G^R (E) = [E + i \eta - H_0 - \Sigma (E)]^{-1}$, where $H_0$ and $\Sigma$ represent the bare single particle Hamiltonian and the impact of non-Hermiticity on the system, respectively~\cite{kozii2017non}. Further, $\Sigma$ can characterize the self-energy in a many-body scenario where the quasi-particles have a finite lifetime~\cite{kozii2017non}. Additionally, it can capture the presence of gain and loss terms in an effective non-Hermitian Hamiltonian, particularly when describing an open system coupled to a bath with appropriate time dynamics~\cite{carmichael2009open}. 

% The Green's function for a non-Hermitian Hamiltonian (away from exceptional points) can be expressed as

% \begin{equation}
% G^R (E) = \sum_n \dfrac{|\psi^{R}_n \rangle \langle \psi^{L}_n |}{E + i \eta - E_n},
% \label{eq:green2}
% \end{equation}

% where $|\psi^{R}_n\rangle$ and $|\psi^{L}_n\rangle$ refer to the right and left eigenstates of the non-Hermitian Hamiltonian, respectively, i.e., $H| \psi^{R}_n \rangle= E_n |\psi^{R}_n \rangle$ and $H^{\dagger}| \psi^{L}_n \rangle= E^{*}_n |\psi^{L}_n \rangle$.Note that the left and right eigenstates constitute a bi-orthogonal basis $\langle \psi_m^{L}|\psi_n^{R}\rangle=\delta_{mn}$~\cite{brody2013biorthogonal,bergholtz2021exceptional,kunst2018biorthogonal}.

%%%%%%%%%%%%%%%%%%%%%%%%%%%%%%%%%%%%%%%%%%%
\begin{figure}	
\includegraphics[width=0.96\linewidth]{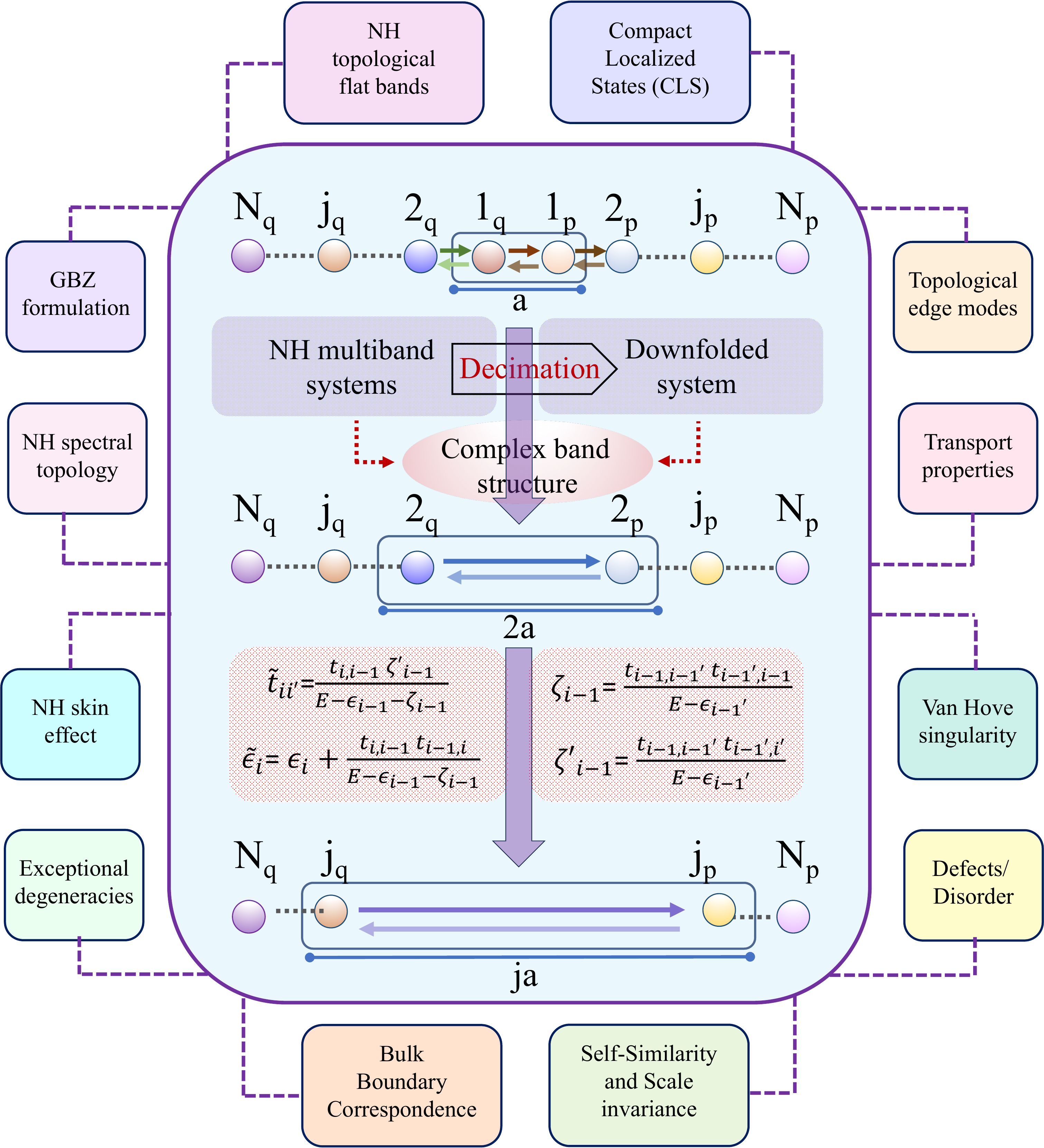}
\caption{\textbf{Application of the decimation scheme in non-Hermitian physics.} Schematic representation of the real space decimation scheme to down-fold a non-Hermitian Hamiltonian. Initially, different blocks of lattice sites ($n_{p/q}$, where $n = 1,2,3,...N$) are interconnected by non-reciprocal hopping matrices marked by solid and dashed arrows. Each block can include both gain and loss terms and non-reciprocal intra-block hopping matrices. Each decimation step invariably reduces the number of blocks but at the expense of renormalized tight-binding parameters. The modified parameters are also shown for a general $i$-th step decimation process. The down-folded lattice precisely mimics both the real and imaginary spectra of the original non-Hermitian model. The potential application of this technique in revealing numerous properties of a non-Hermitian system are also highlighted.}
\label{schematic-decimation}	
 \end{figure}
%%%%%%%%%%%%%%%%%%%%%%%%%%%%%%%%%%%%%%%%%%%

For a lattice model, a preferred subset of ``degrees of freedom" (or variables) can be eliminated from the original set of linear equations for the Green's function. In principle, the decimation process can be an iterative process that systematically coarse-grains the parameter space; however, it yields the exact full density of states (DOS)~\cite{ashraff1988exact}. In a similar vein, here, we shall explore how the degree of complexity of any non-Hermitian multiband system can be substantially reduced by downfolding the corresponding Hamiltonian matrix. For a comprehensive understanding of the approach, let us consider a linear chain comprising a total of $N_p + N_q$ blocks, as depicted in Fig.~\ref{schematic-decimation}, containing information about the non-reciprocal hopping and on-site gain and loss terms. These blocks can correspond to either a single site or a collection of sites for which the tight-binding analogue of the Schr\"{o}dinger equation
can be written as

\begin{equation}
    (E-\epsilon_i)\phi_i=\sum_{j} t_{ij} \phi_j,
    \label{eq:rsrg1}
\end{equation}

where $E \: \mathbb{I}$, $\epsilon_i$, $\phi_i$ and $t_{ij}$ matrices represent the eigenenergy, onsite potential and probability amplitude at the $i$th block, and hopping parameter between $i$-th and $j$-th block, respectively. The above strategy essentially reduces the order of the Hamiltonian matrix after each decimation step keeping the characteristic equation invariant. The transformed tight-binding parameters ($\tilde{\epsilon}_{k}$, $\tilde{t}_{kl}$) at any given decimation step contain all the information of the previous step ($\epsilon_{i}$, $t_{ij}$) as given below,

\begin{align}
       \tilde{\epsilon}_{i} &= \epsilon_{i} + \frac{t_{i,(i-1)} t_{(i-1),i}}{E-\epsilon_{(i-1)}  - \frac{t_{(i-1),(i-1)^{\prime}} t_{(i-1)^{\prime},(i-1)}}{E-\epsilon_{(i-1)^{\prime}}}}, \nonumber \\
        \tilde{t}_{i,i^{\prime}} & = \frac{t_{i,(i-1)} \: t_{(i-1),(i-1)^{\prime}} \: t_{(i-1)^{\prime},i^{\prime}}}{\left(E-\epsilon_{(i-1)^{\prime}}\right) \left[E - \epsilon_{(i-1)} - \frac{ t _{(i-1),(i-1)^{\prime}}  t _{(i-1)^{\prime},(i-1)}}{E-\epsilon_{(i-1)^{\prime}}} \right] }.
        \label{eq:onshop}
\end{align}

In the above Eq.~\ref{eq:onshop}, the $p$ and $q$-sites are represented by primed and unprimed parameters and should be interchanged for $p \leftrightarrow q$. This decimation scheme is an iterative process, and the non-Hermitian Green's function can readily be calculated from the renormalized Hamiltonian at each decimation step. In other words, the entire system information can be encoded into an effective two-site problem containing decimated onsite parameters of $N_{p}$ and $N_{q}$ and hopping parameters between the same. However, in this case, the most pertinent question remains -- Can the real space decimation scheme capture the complete complex multiband topology? Next, we present a comprehensive resolution to this question at hand.

\section{Non-Hermitian band topology through decimation}

\begin{figure}
 \includegraphics[scale=0.38]{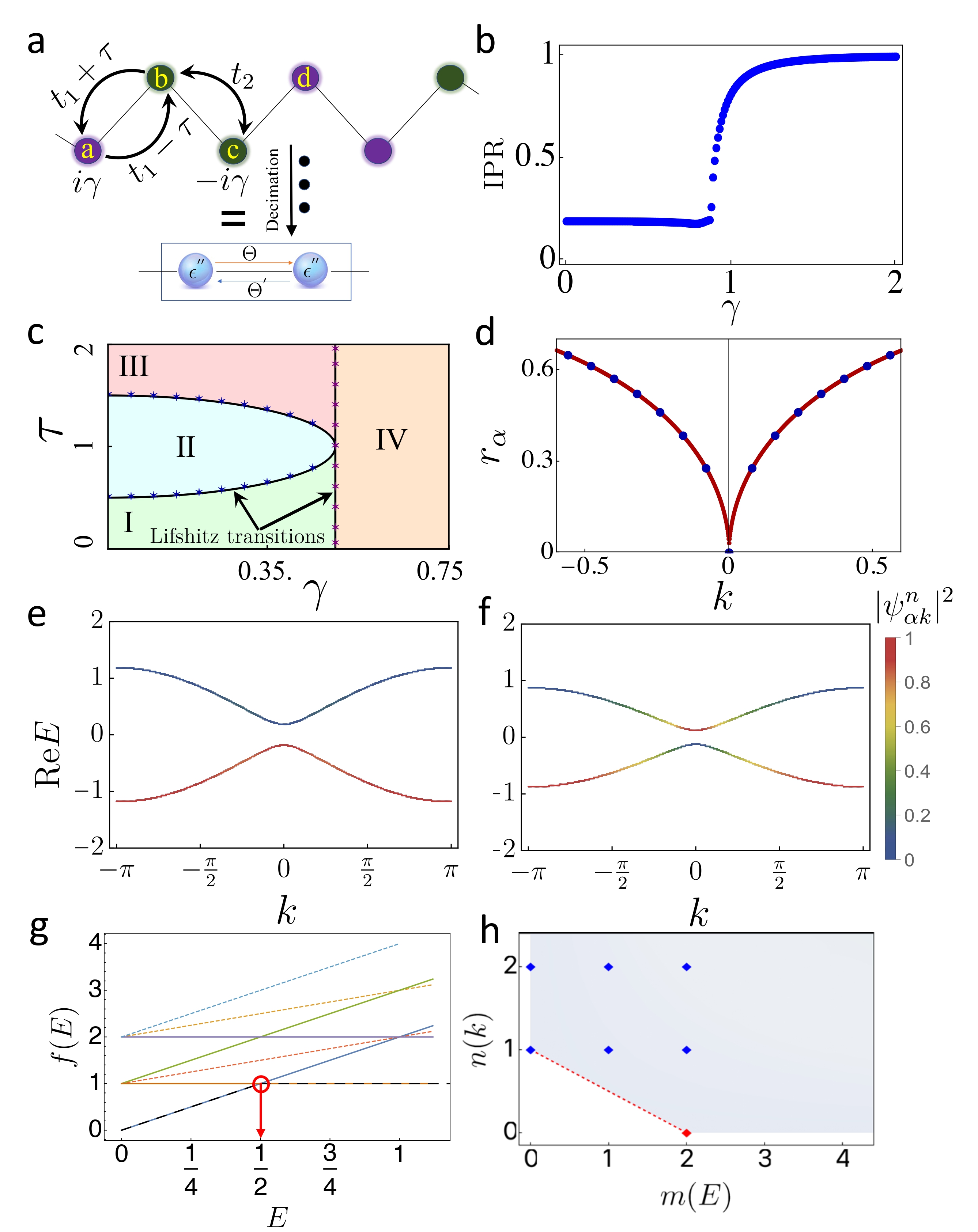}
\caption{\textbf{Analysis of NH spectral topology in a generalized SSH chain through decimation scheme.} (a) Illustration of the decimation scheme for the generalized SSH chain (see Eq.~\ref{eqn:Ham}) to obtain a downfolded two-site model which encodes the essential topological information of the NH spectral topology of the original model. (b) Phase diagram of the model illustrating four phases undergoing Lifshitz transition with distinct point and line gap topology. The phase boundaries are the locus of the exceptional points characterized by phase rigidity in (c). The comparison of phase rigidity obtained from the decimated model (blue dots) and the parent model (solid red line) is also shown, which reveals a striking resemblance. At the $\tau=0$ limit, the system encounters an SSH-like (Hermitian) topological phase transition governed by real energy gap closing. The topological real energy zero modes under the low energy approximation can be traced out by IPR from the decimated model (d). Furthermore, the orbital character evaluated from the decimated model demarcates the trivial (e) and non-trivial phases (f) of the system with a characteristic twist around $k=0$, which further reveals the SSH-like phase transitions with the gap closing at $\gamma_{c}=\sqrt{3}/2$. We characterize the second-order exceptional points along the phase transition lines using (g) tropical geometric framework and (h) Newton polygon formalism discussed in the supplementary material~\cite{supplement}. In the tropical geometric framework, the bend locus of tropicalization indicates the square root dispersion, whereas the Newton polygon formalism illustrates the same through its negative slope shown in the red dotted line. We set $t_1=0.5$ and $t_2=1.0$.} \label{ssh-model}
\end{figure}

To answer the aforementioned question, we consider a non-Hermitian four-band model featuring both non-reciprocal hopping and inversion symmetric imaginary potentials, as illustrated in Fig.~\ref{ssh-model}. The Hamiltonian is $H=H_{hop}+H_{pot}$, where the individual terms are given by~\cite{lieu2018topological}

\begin{eqnarray}
H_{hop} =  - \sum_{j} [t_1 (c^\dagger_{j,A} c_{j,B} + h.c.) +
t_2 (c^\dagger_{j+1,A} c_{j,B} + h.c.)]  \notag \\ 
+ \sum_{j} \tau (c^\dagger_{j,B}c_{j,A} - c^\dagger_{j,A}c_{j,B}), \notag \\
H_{pot} =  i\gamma  \sum_{j} (c^{\dagger}_{2j-1,A}c_{2j-1,A} +c^\dagger_{2j,B}c_{2j,B}  - c^\dagger_{2j,A}c_{2j,A} \hspace{0.28cm}
\notag \\ -c^{\dagger}_{2j-1,B}c_{2j-1,B} ), \hspace{0.6cm}
\label{eqn:Ham}
\end{eqnarray}

where $c^{\dagger}_{j,\alpha} (c_{j,\alpha})$ are the fermionic creation (annihilation) operators for the sublattice $\alpha=A, B$, and the lattice site is indexed by $j$ [see Fig.~\ref{ssh-model} (a)]. The term $H_{hop}$ with non-reciprocal intra-unit cell hopping $t_1 \pm \tau$ and inter-unit cell coupling $t_2$ describes the two sublattices in the non-Hermitian Su-Schrieffer-Heeger (SSH) model. Additionally, we introduce $H_{pot}$, i.e., the imaginary staggered potential that respects inversion symmetry. The system Hamiltonian of the lattice in Bloch space is obtained as

\begin{equation}
  H_{k}=  \begin{pmatrix}
    i\gamma & t_1-\tau & 0 & t_2 e^{-i k} \\
   t_1+\tau & -i\gamma & t_2 & 0 \\
    0 &  t_2 & -i\gamma & t_1-\tau \\
    t_2 e^{i k} & 0 &  t_1+\tau & i\gamma \\
    \end{pmatrix}.
\end{equation}

The selection of the generalized non-Hermitian SSH model as our focus is particularly appropriate, considering the extensive research dedicated to both $H_{hop}$ and $H_{pot}$ individually. This choice allows us to leverage the substantial body of knowledge on these models~\cite{herviou2019defining,wu2021topology,jin2019bulk,kunst2018biorthogonal,song2019non,esaki2011edge,alvarez2018non,borgnia2020non} and explore its intriguing physics within the context of the RSRG decimation scheme. The system described by the Hamiltonian, $H=H_{hop}+H_{pot}$, has a discrete translational symmetry. Furthermore, the Bloch Hamiltonian, $H_k$, in the momentum space respects the ramified particle-hole symmetry ($\text{PHS}^{\dagger})$ denoted by the unitary matrix $\hat{\mathcal{S_{-}}}$~\cite{kawabata2019symmetry}. The $\text{PHS}^{\dagger}$ operator is defined by: $\hat{\mathcal{S_{-}}}H_{k}^{*}$$\hat{\mathcal{S}}$$^{-1}_{-}=-H_{-k}$, where $\hat{\mathcal{S_{-}}}= \sigma_{0}\otimes \sigma_{z}$.

We move on to the real space and employ the decimation scheme to decimate the four-band model to obtain an effective two-band model with renormalized energy-dependent coupling and onsite (gain/loss) terms. We will show that this two-band model can precisely mimic the original system, capturing all its essential physics, and also decipher the known phase diagram for $\tau=0$ as a special case (see supplement for details~\cite{supplement}). Next, we present a systematic study of the decimated model and the underlying non-Hermitian phase transitions as a function of non-reciprocity, $\tau$, and gain-and-loss co-efficient, $\gamma$. 

We have judiciously integrated out the middle two (green) sites from the original lattice [see Fig.~\ref{ssh-model} (a)]. In particular, we have obtained the following coupled equations, $[E-\epsilon^{''}]\phi_a =  \tau \phi_d +t_2 \phi_d^{'}$ and $ [E-\epsilon^{''}]\phi_d =  \tau \phi_a+t_2 \phi_a^{'}$. This yields an effective two-band model with energy-dependent renormalized couplings and gain and loss terms. 

\begin{equation}
H_d=
\begin{pmatrix}
  \epsilon^{''} & \Theta + t_2e^{-ika}\\ 
  \Theta^{'} + t_2e^{ika} &  \epsilon^{''}
\end{pmatrix},
\label{four-cite-decimated}
\end{equation}

where $\epsilon^{''}=i\gamma+(t_1-\tau)(t_1+\tau)/(E-\epsilon^{'})$, $\epsilon^{'}=-i\gamma+t_{2}^{2}/(E+i\gamma)$, $\Theta={(t{_1}-\tau)^{2}t_2}/[(E+i\gamma)(E-\epsilon^{'})]$ and $\Theta^{'}={(t{_1}+\tau)^{2}t_2}/[{(E+i\gamma)(E-\epsilon^{'})}]$. The phase diagram as a function of $\tau$ and $\gamma$ and the concomitant band dispersion corresponding to each phase for this effective two-band model is shown in Fig.~\ref{ssh-model} (c) (see also supplement~\cite{supplement}). The dispersion exactly matches the original four-band model. This alignment persists across the entire parameter range, showcasing notable phenomena such as the presence of exceptional points and topological phase transitions which we discuss next.

The single particle spectrum shows gap closings for absolute values of energies for lines $\pm t_2 \pm \sqrt{t_1^2 - \gamma^2} $ as well as for real energies $t_1=\gamma$ dividing the phase diagram into four regions [see Fig.~\ref{ssh-model}(b)]. We discuss various phases and their transitions as well as characterize the spectral topology using the non-Hermitian winding number in corroboration with the non-decaying chiral modes (for details see supplement~\cite{supplement}). 

The exceptional physics occurring around the phase transition lines can be effectively characterized through a tropical geometric structure~\cite{banerjee2023tropical} and Newton Polygons~\cite{jaiswal2023characterizing}, despite the presence of energy-dependent tight-binding parameters [see Fig.~\ref{ssh-model} (g) and (h) for the illustration and supplement for more details~\cite{supplement}]. To fully characterize the eigenfunctions using the decimated model, it is essential to solve the following eigenvalue equation,
$H_d(t_1,t_2,\gamma,\tau,E) \psi_d=E_d (t_1,t_2,\gamma,\tau,E)\psi_d$. We solve for the eigenvalues and eigenstates numerically. This process entails assuming different trial values of complex energy and subsequently checking which of these satisfies the characteristic equation. By systematically iterating through these trials, we are able to identify the complex energy values that correspond to the eigenvalues of interest. Thereby, we also obtain the eigenfunctions for our decimated system. In Fig.~\ref{ssh-model}(c), we show the scaling of the phase rigidity, $r_{d} =\dfrac{\langle \psi_{d}^L|\psi_{d}^R\rangle}{\langle \psi_{d}^R|\psi_{d}^R\rangle}$~\cite{eleuch2017resonances}, near a phase transition line. The phase rigidity which is solely a function of the eigenvectors, correctly reproduces the behaviour of our original model [see Fig.~\ref{ssh-model} (c) for a comparison with the parent model]. This further confirms the existence of an exceptional locus featuring the coalescence of eigenvectors.

At this point, it is interesting to note that in the limit $\tau \rightarrow 0$, the model corresponds to the $\text{BDI}^{\dagger}$ class of the 38-fold topological classification of non-Hermitian systems, which in turn suggests that the topological phase transition of the system is governed by the closure of the real part of the energy band gaps~\cite{kawabata2019symmetry,wu2021topology}. To corroborate this, we analyse the low energy limit of the decimated model and demonstrate the orbital character associated with it. The absence (presence) of characteristic twists around $k=0$ confirms the distinction of the trivial (non-trivial) phase, as well as the existence of a critical point $(\gamma_c)$ where the real energy gap closes. The closing of energy at this critical point signifies a topological phase transition [See Fig.~\ref{ssh-model}(e) and (f); for more details, see Supplement~\cite{supplement}]. The inverse participation ratio, IPR= $\sum_{\alpha}|\psi_\alpha(x)|^4/\big(\sum_{\alpha}|\psi_\alpha(x)|^2\big)^2$, based on eigenfunction characteristics which quantitatively measures state localization under open boundary conditions, allows us to identify and analyze the presence of localized edge modes within the system, as shown in Fig.~\ref{ssh-model}(d). Non-Hermitian systems exhibit remarkable sensitivity to boundary conditions, resulting in a significant disparity between spectra under periodic (PBC) and open (OBC) boundary conditions. This broken BBC challenges conventional understanding within the Hermitian paradigm~\cite{xiong2018does,xiao2020non,helbig2020generalized}.

\begin{figure}	
\includegraphics[width=0.97\linewidth]{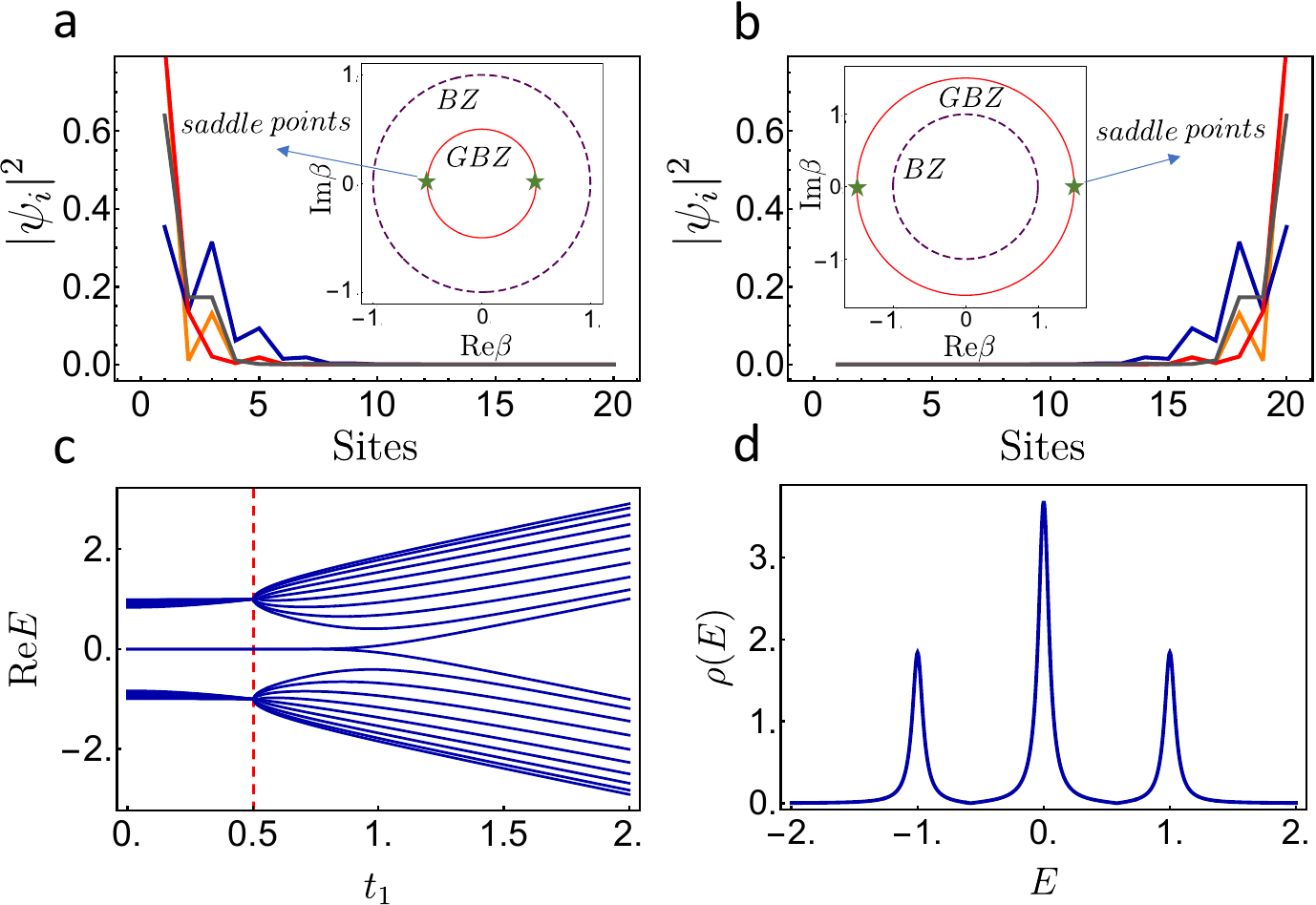}
\caption{\textbf{Demonstration of chiral skin effect, bulk boundary correspondence and van Hove singularity through GBZ formulation within decimation scheme.} Localization of bulk wavefunctions at (a) Left $(\tau<0)$ and (b) right $(\tau>0)$ edge illustrating the NH skin effect. The insets show GBZ (red) and BZ (dotted black), depicting a non-unit and unit circle, respectively, in the complex plane. The radius of the GBZ greater (lesser) than unity indicates the right (left) localized states and skin effect. The GBZ becomes a unit circle with $\tau=0$, the system restores conventional bulk boundary correspondence. (c) $t_1=\tau$ is the locus of the $PT$-symmetry breaking point with a higher order exceptional point. At this critical parameter value, the two saddle points (indicated by blue arrows) in the GBZ coalesce. Consequently, a van Hove singularity is observed in the DOS as a function of real energy.}\label{VanHove Singularity}	
\end{figure}

To investigate the effect of boundary conditions on the BBC, we consider the one-dimensional real-space tight-binding chain corresponding to the momentum-space Hamiltonian in Eq.~\ref{four-cite-decimated}. Next, we invoke OBC and evaluate the transfer matrix $(\mathcal{T})$ through the singular value decomposition of the hopping matrix~\cite{kunst2019non}, and in terms of the onsite Green's function (see supplement for details~\cite{supplement}). The transfer matrix approach smoothly connects the OBC and PBC regime and characterizes the BBC through the unimodularity condition, $\text{det} \mathcal{T}=1$. The corresponding analysis of our decimated model gives rise to $\text{det} \mathcal{T}=\frac{ (t_1+\tau)^2}{ (t_1-\tau)^2}$ (see supplement for the details of the calculation~\cite{supplement}), which immediately suggests that in the reciprocal hopping limit $(\tau \rightarrow 0)$, the BBC is restored, with identical OBC and PBC spectra. In contrast, the non-reciprocal hopping limit leads to a non-unimodular condition, indicating a disparity between OBC and PBC spectra, thus invalidating the BBC. These results align with the generalized transfer matrix approach introduced by Kunst \emph{et al.}~\cite{kunst2019non} and correspond to a special case of their method. At the vanishing determinant condition, $\text{det}\mathcal{T}=0$, the real space spectrum gives rise to higher-order exceptional points with an algebraic multiplicity commensurate with the system size, whereas the geometric multiplicity remains one indicating the presence of NHSE.

To delve deeper into the physics of the skin effect in the decimated system, we adopt the non-Bloch theory with the GBZ scheme~\cite{yao2018edge,yokomizo2019non}. Here, the conventional Bloch phase factor, $e^{i k}$, is replaced by $\beta = |\beta| e^{i k}$ in the Hamiltonian, enabling a direct mapping between non-Bloch topology and open boundary spectra. This framework based on the decimated model also captures the NHSE~\cite{yokomizo2019non,yang2020non}. The real-space eigen-equation corresponding to Eq.~\ref{four-cite-decimated} leads to the conditions, $\Theta^{'} \psi_{a,n}+t_2 \psi_{a,n+1}+\epsilon^{''}\psi_{b,n}=E \psi_{b,n}$ and $\Theta \psi_{b,n}+t_2 \psi_{b,n-1}+\epsilon^{''}\psi_{a,n}=E \psi_{a,n}$. Analogous to Ref.~\cite{yokomizo2019non}, we consider the ansatz governed by the spatial periodicity of the system $(\psi_{a,n},\psi_{b,n})=\beta^n (\psi_{a},\psi_{b})$ and we obtain the coupled equations $\Theta^{'}\psi_a+t_2 \beta \psi_a +(\epsilon^{''}-E)\psi_b=0$ and $\Theta \psi_b+ t_2 \beta^{-1}\psi_b+(\epsilon^{''}-E)\psi_a=0$. This leads to the condition 

\begin{equation}
g(E, \beta)=\beta^2 t_2 \Theta + \beta \big\{\Theta^{'}\Theta + t_2^2-(\epsilon^{''}-E)^2 \big\}+t_2 \Theta^{'}=0,
\label{beta-equn}
\end{equation}

from which $\beta$ can be estimated with two solutions, $\beta_1$ and $\beta_2$, which satisfy $\beta_1 \beta_2=\Theta^{'}/\Theta$. It can be shown that the bulk states of a long
chain demand the condition, $|\beta_1|=|\beta_2|$, which eventually leads to the solution $|\beta|=|\beta_1|=|\beta_2|=\sqrt{|\Theta^{'}/\Theta|}=\sqrt{| (t_1+\tau)^2/ (t_1-\tau)^2}|$. Furthermore, $|\beta|<1 (|\beta|>1)$ corresponds to states localized at the left (right) end of the chain. We refer to Fig.~\ref{VanHove Singularity} (a) and (b) for an illustration of both conditions with parameter dependence $\tau<0$ $(\tau>0)$ exhibiting the left (right) localized bulk modes. Interestingly, the occurrence of the skin effect in the different phases is evinced by the finite spectral area (closed curves) in the complex energy plane of the decimated model under PBC (see also supplement~\cite{supplement})~\cite{zhang2020correspondence}. 

\begin{figure}	
\includegraphics[width=0.97\linewidth]{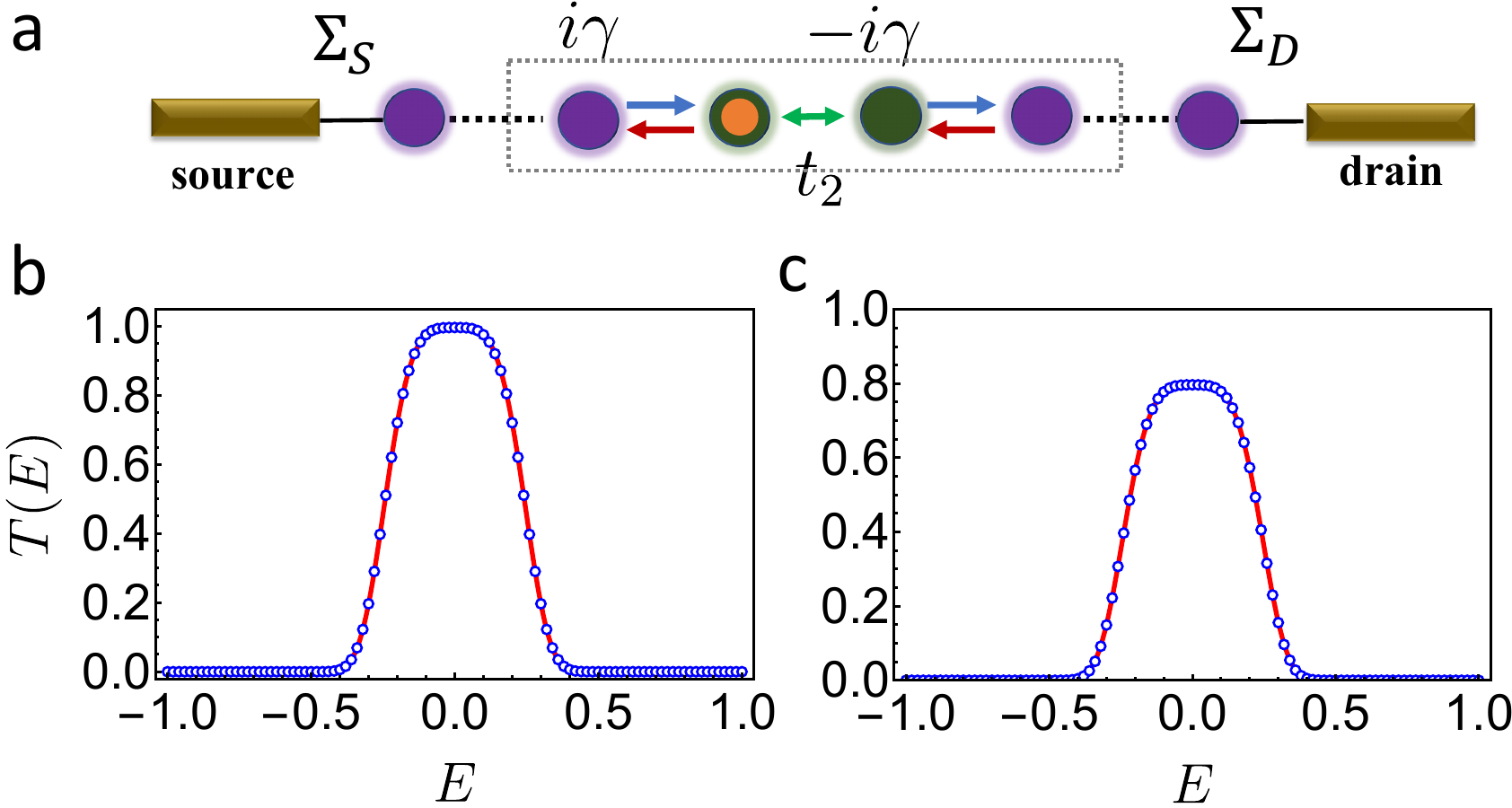}
\caption{\textbf{Transmission probability for original and decimated systems in presence of a defect.} (a) Schematic of the model for calculating the transmission probability of the non-Hermitian four-site chain. The blue (red) arrow indicates the non-reciprocal hopping $t_1\pm\gamma$ along the right (left) direction. Ideal lossless metallic leads are used as electrodes, with source and drain with self-energy matrices $\Sigma_{S}$ and $\Sigma_{D}$, respectively. The orange solid circle represents the defect in the system. The transmission probability of the system (b) without and (c) with defect, when the system is tuned near the exceptional points. The presence of the defect invariably suppresses the transmission probability. The probability obtained from the decimated scheme (shown with blue dots) exactly matches the original model (red solid line). Here we set $t_1 = 0.5$, $t_2 =1.0$, $\gamma = 0.5$, $\tau = -1.0$, and $\delta_d = 1$.} \label{fig:transmission}	
\end{figure}

Next, we discuss zero-mode solutions in the low-energy approximation $(E=\epsilon^{''})$, when Eq.~\ref{beta-equn} leads to the solutions $\beta_{1,2}= -\Theta^{'}/t_2$,   $-t_2/\Theta^{'}$. The zero-energy solutions, where the bulk bands touch the zero energy, can be obtained by equating $|\beta|=|\beta_{1,2}|$, leading to the condition $t_2^2=t_1^2-\tau^2$. We can thereby determine the GBZ, which gives rise to continuum bands under the condition $|\beta_1|=|\beta_2|$ with open boundaries. Considering Eq.~\ref{beta-equn} as the kernel of non-Bloch band theory, the continuum bands can be given by solution to $g(E,\beta)$ with $z_{\text{GBZ}}=\beta= |\beta| e^{i k/2}$ and $k \in [0, 2\pi)$ (see supplement~\cite{supplement}). The GBZ in the complex plane describes the OBC spectrum in non-Hermitian systems with complex deformation of the momentum $k\rightarrow k + i \kappa$. Further, $\kappa= \ln{|\beta|}$ indicates the inverse localization length of skin modes. We next calculate the saddle points and their energies which satisfy the condition $g(E,\beta)=0$ and $\partial_{\beta} g(E,\beta)=0$ simultaneously (see supplement for details~\cite{supplement}). The saddle points $(\beta_{s(1,2)}=\pm (t_1+\tau)/(t_1-\tau))$, and their coalescence leads to $t_1= |\tau|$ and $\text{det}\mathcal{T}=0$, which in turn, gives rise to interesting consequences~\cite{hu2022geometric}, which we explore in terms of the decimated model. 

In the present scenario, $\text{det}\mathcal{T}=0$ indicates a higher order exceptional point encountering a $PT$ (parity-time) symmetry transition point where the purely real energy spectra bifurcate in the complex plane~\cite{bender1998real,el2018non}. Interestingly, the merging of saddle points corroborated by the $PT$ transition is stipulated by a singularity in the DOS along the real axis. To substantiate this, as a byproduct of our formalism, we calculate the DOS in terms of Green's function at $t_1=\tau$, which is the $PT$ breaking point [see Fig.~\ref{VanHove Singularity}(c)]. The saddle point coalescence on the GBZ manifests a divergence in the DOS [see Fig.~\ref{VanHove Singularity}(d)], leading to a non-Bloch van Hove singularity~\cite{hu2022geometric}. 

Furthermore, we have investigated the transmission probability, $T(E)$, through the non-Hermitian system placed between two ideal metallic electrodes, i.e., source and drain. The transmission probability can be evaluated as $T(E) = \mathrm{Tr}[\Gamma_{S} G(E) \Gamma_{D} G(E)^{\dagger}$] within the Landauer-Buttiker formalism~\cite{datta1997electronic,li2021non}. In the above equation, $\Gamma_{S}$ and $\Gamma_{D}$ can be estimated from the self-energy matrices $\Gamma_{S(D)} = i \left[ \Sigma_{S(D)} - \Sigma^{\dagger}_{S(D)} \right]$ and therefore depend on the coupling between the electrodes and the system. In particular, we have considered the above-mentioned non-Hermitian chain having a total of $2n+1$ unit cells (here, $n=50$) and the parameters are so chosen that we reside near an exceptional point, i.e., $\tau = -t_2 \pm \sqrt{t_1^2 - \gamma^2}$, as schematically represented in Fig.~\ref{fig:transmission}(a). It is fascinating to note that the transmission probability of the original system is precisely in line with that of the decimated system [Fig.~\ref{fig:transmission}(b)]. The nature of the transmission probability near the exceptional point agrees well with the previous reports~\cite{li2019exceptional,soori2022transmission}.

Needless to mention, the downfolding of the Hamiltonian is indeed an iterative process, and the transmission probability will be exactly the same for any decimated system size, even in the presence of any defect. To establish this, we have incorporated a defect with onsite energy $\delta_d$, which could be present at any random lattice site. Let us consider the defect site is in the $(n+1)$-th unit cell of the system with $2n+1$ lattice sites. As we expect, the transmission probability reduces with the strength of the defect potential, which is entirely captured by both the full and decimated Hamiltonians as shown in Fig.~\ref{fig:transmission}(c). In the present case, the lattice sites of the remaining $2n$ unit cells have been decimated to half; however, the same methodology can be applied iteratively to reduce the lattice sites further, similar to the above case. We discussed the central features of non-Hermitian topological systems in light of the decimation scheme, and the crux of the analysis is that the complete non-Hermitian spectral topology can be encapsulated in the downfolded version of an extensive system.

\section{Non-Hermitian flat band through decimation} 

\begin{figure}	
\includegraphics[width=0.96\linewidth]{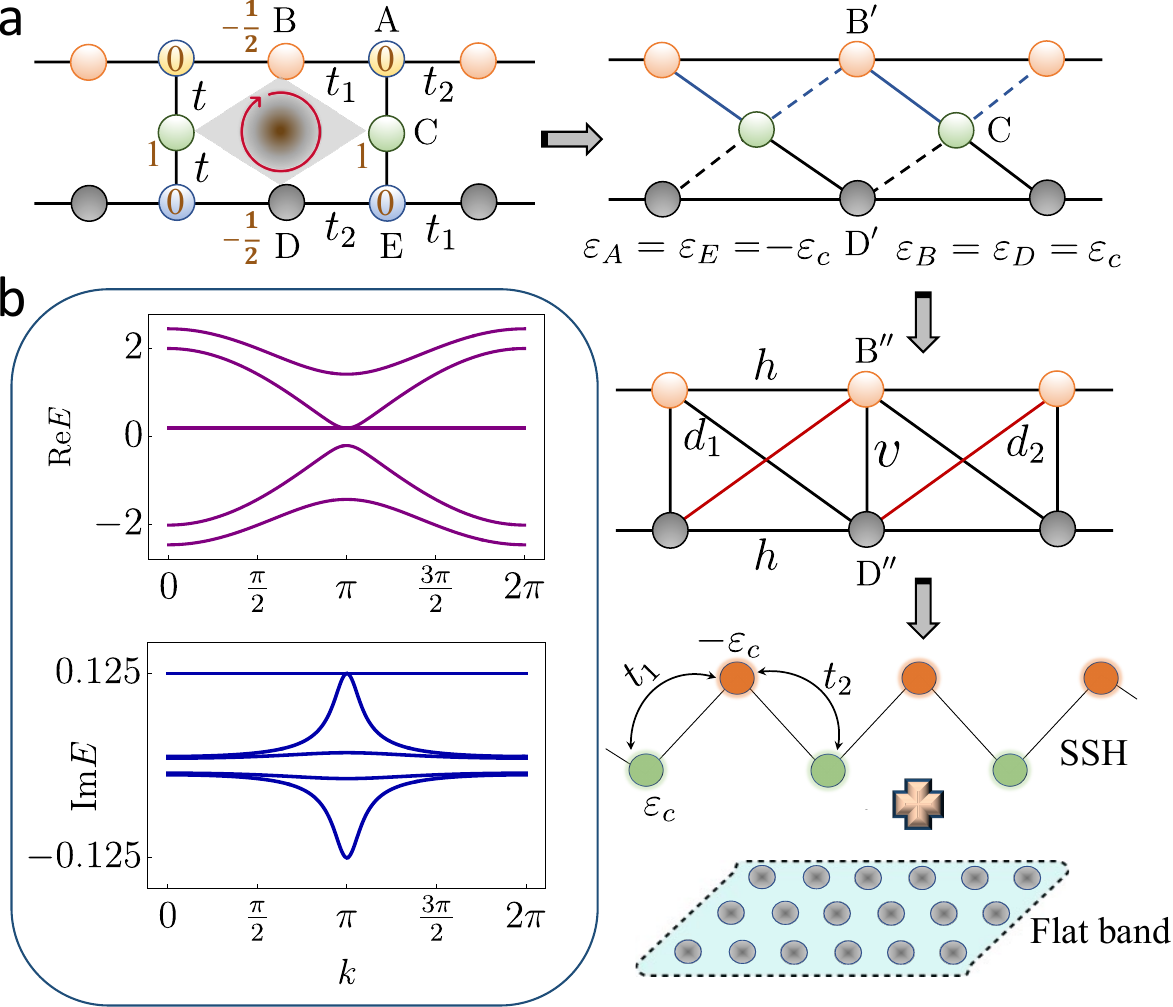}
\caption{\textbf{Non-Hermitian flat bands through decimation.} (a) Quasi one-dimensional Lieb lattice model with two triply coordinated ($A$ and $E$) and three doubly coordinated ($B$, $C$, and $D$) sites. The same sites of the nearest unit cells are denoted by subscripts $l$ and $r$ for the left and right directions, respectively. Non-Hermiticity in the model has been incorporated through onsite gain and loss following the conditions $\varepsilon_A=-\varepsilon_B=-\varepsilon_c=-\varepsilon_D=\varepsilon_E$. The real space decimation process reduces the lattice to an equivalent two-level non-Hermitian ladder system in two steps. First, the triply coordinated sites $A$ and $E$ are decimated, followed by the elimination of the $C$ site. Finally, the ladder network is decoupled into a massive non-Hermitian SSH chain that carries the information about the topological band inversion and an array of isolated lattice sites with localized orbitals that constitute a flat band at $E=\varepsilon_c$. (b) The emergence of a flat band at a complex energy value ($0.2+ 0.125i$) along with the massive non-Hermitian SSH-like dispersive band structure. The corresponding compact localized states are indicated by the gray regions in the original lattice. Note that the localization of electronic states is caused by the zero probability amplitude of the atomic sites $A$ and $E$. The parameters used here are $\varepsilon_c=0.2+ 0.125i$ and $t_1=t_2=t=1.0$.}
\label{five-site-model}	
\end{figure}

Now, we discuss how the real space decimation technique can describe the fascinating flat band physics and emergence of compact localized states (CLSs) in non-Hermitian lattice models. As a test bed, we consider a quasi one-dimensional Lieb lattice with five sites per unit cell and incorporate non-Hermiticity through onsite gain and loss~\cite{vicencio2015observation,whittaker2018exciton,liu2020universal}. We note that the downfolding approach presented here can be efficiently extended to any other lattice type with a flat band, where the non-Hermiticity can be of any kind, including nonreciprocal hopping. One of the realistic pathways to achieve such a non-Hermitian Lieb (nH-Lieb) lattice is to construct a photonic crystal by employing periodically arranged evanescently coupled waveguides~\cite{xia2021higher}. The Hamiltonian for this five-site nH-Lieb lattice model can be written as follows

\begin{eqnarray}
\mathcal{H}= \sum_{\nu=\{a,b,c,d,e\}} \varepsilon_\nu \nu_n^{\dagger}\nu_n +
\sum_{n}[ a_n^{\dagger}(\tilde{t_1} b_n+ \tilde{t_2} b_{n-1}+ t c_n)  + \nonumber \\
e_n^{\dagger}(\tilde{t_2} d_n+ \tilde{t_1} d_{n-1}+t c_n)+ h.c.], \hspace{0.6cm}
\label{eq:nhLieb}
\end{eqnarray}

where $a_n^{\dagger} (a_n)$, $b_n^{\dagger} (b_n)$, $c_n^{\dagger} (c_n)$, $d_n^{\dagger} (d_n)$ and $e_n^{\dagger} (e_n)$ are the fermionic creation (annihilation) operator at $n$-th unit cell for five distinct sublattices $A$, $B$, $C$, $D$ and $E$ [see Fig.~\ref{five-site-model} (a)]. The parameters $t,t_1$ and $t_2$ are the coupling coefficients between different neighbouring sites. Note that the lattice sites $A$ and $E$ are triply coordinated while the rest ($B$, $C$, and $D$) have coordination number two. Similar to the previous discussion, here, we aim to reduce the degree of complexity by decimating the five-site lattice to an equivalent two-level problem. For this purpose, we have first eliminated the triply coordinated sites $A$ and $E$ by substituting the corresponding eigenvectors $\phi_a$ and $\phi_e$ using

\begin{equation}
    (E-\varepsilon_{a/e})\phi_{a/e}  = t_1 \phi_{b/dr} + t_2 \phi_{br/d}+t\phi_{c},
\end{equation}    

into the following expressions

\begin{align}
     (E-\varepsilon_c)\phi_c & = t \phi_a + t \phi^{}{_e}, \nonumber \\
     (E-\varepsilon_{b/b/d/d})\phi_{b/br/d/dr} & = t_1 \phi_{a/ar/el/e} + t_2 \phi_{al/a/e/er}.
\end{align}

The subscripts $l$ and $r$ signify the sites of the nearest left and right unit cell, respectively. Fig.~\ref{five-site-model} (a) illustrates that this decimation process leads to an effective three-level lattice comprising of $B$, $C$, and $D$ sites with renormalized hopping parameters given as $\lambda_{a_1} = t t_1/(E-\varepsilon_a)$, $\lambda_{a_2}= t t_2/(E-\varepsilon_a)$, $\lambda_{e_1}= t t_1/(E-\varepsilon_e)$, $\lambda_{e_2}= t t_2/(E-\varepsilon_e)$, $\lambda_{a_{12}}= t_1 t_2/(E-\varepsilon_a)$, and $\lambda_{e_{12}}=t_1 t_2/(E-\varepsilon_e)$. In addition, we also need to consider the renormalized values for the onsite energies of $C$, $B$, and $D$, which are $\varepsilon_{c}^{\prime}=\varepsilon_c+t^2/(E-\varepsilon_a)+t^2/(E-\varepsilon_e)$, $\varepsilon_{b}^{\prime}=\varepsilon_b+(t_{1}^{2}+t_{2}^{2})/(E-\varepsilon_a)$, and $\varepsilon_{d}^{\prime}=\varepsilon_d+(t_{1}^{2}+t_{2}^{2})/(E-\varepsilon_e)$. It is evident that, as per our expectation, the new set of renormalized hopping integrals and onsite potentials depend on the decimated parameters of the original lattice. Additionally, the energy dependency of the tight-binding parameters protects the order of the characteristic equation. Now, one can execute another decimation step for further renormalizing the three-site lattice with energy-dependent parameters to an equivalent two-level system. The convenient approach, in this regard, is to choose a scheme that dissolves the information carried by the wavefunction $\phi_c$ into the ladder like network made up of $B$ and $D$ sites [Fig.~\ref{five-site-model} (a)]. The final renormalized lattice contains only two types of onsite energy terms and five distinct hopping parameters given by 

\begin{align}
    \varepsilon^{\prime \prime}_{b} & = \varepsilon^{\prime}_{b} + (\lambda_{a_1}^{2}+\lambda_{a_2}^{2})/(E-\varepsilon_{c}^{\prime}), \nonumber \\ \varepsilon^{\prime \prime}_{d} & = \varepsilon^{\prime}_{d} + (\lambda_{e_1}^{2}+\lambda_{e_2}^{2})/(E-\varepsilon_{c}^{\prime}), \nonumber \\
    h_1 & = \lambda_{a_{12}}+(\lambda_{a_{1}} \lambda_{a_{2}})/(E-\varepsilon_{c}^{\prime}), \nonumber \\
    h_2 & = \lambda_{e_{12}}+(\lambda_{e_{1}} \lambda_{e_{2}})/(E-\varepsilon_{c}^{\prime}), \nonumber \\
    d_1 & = \lambda_{a_{1}}\lambda_{e_{1}}/(E-\varepsilon_{c}^{\prime}), \nonumber \\
    d_2 & = \lambda_{a_{2}}\lambda_{e_{2}}/(E-\varepsilon_{c}^{\prime}), \nonumber \\
     v & =(\lambda_{a_{1}}\lambda_{e_{2}} + \lambda_{a_{2}}\lambda_{e_{1}})/(E-\varepsilon_{c}^{\prime}).
\end{align}

Therefore, the decimation process downfolds the tight-binding Hamiltonian to a $2 \times 2$ matrix at the expense of allowing new types of energy-dependent hoppings terms and onsite potentials. 

In order to facilitate the understanding of the flat band physics in this nH-Lieb lattice, let us focus on the following case, $\varepsilon_b = \varepsilon_d =\tilde{\varepsilon}$ and $\varepsilon_a = \varepsilon_e = -\tilde{\varepsilon}$ that further leads to the relations $\varepsilon_{b}^{\prime \prime} = \varepsilon_{d}^{\prime \prime} = \Delta$ (say), $h_1 = h_2 = h$, and $v^2=4d_1d_2$. The above relations simplify the eigenvalues of the final Hamiltonian in the following form

\begin{equation}
    E_{\pm} = [\Delta + 2 h \:\textup{cos}(k)] \pm [d_1+d_2+v \: \textup{cos}(k)].
    \label{eq:solutions}
\end{equation}

The above Eq.~\ref{eq:solutions} apparently indicates that the coefficient ($v+2h$) causes the solution $E_{+}$ to be dispersive. However, by substituting all the energy-dependent parameters in the above expression and simplifying, we can show that the effective value of ($v+2h$) is zero. Therefore, the solution is essentially non-dispersive, and the position of the flat band lies at an energy value $E=\Delta+d_1+d_2$. In other words, the particular choice of onsite potentials $\varepsilon_a=-\varepsilon_b=-\varepsilon_d=\varepsilon_e=-\varepsilon_c$ always provides a flat band at the complex energy value $E = \varepsilon_c$, as depicted in Fig.~\ref{five-site-model}(b). This eigenenergy solution is equivalent to that of periodically arranged isolated sites with energy $\varepsilon_c$ where there is no orbital overlap between neighboring sites. The corresponding single-particle real space eigenfunctions constitute the CLS, which can be analytically obtained using our method. In particular, solving the difference equations for a particular eigenvalue that offers a flat band will manifest the probability distribution for the CLS. In the present case of nH-Lieb lattice, probability amplitudes of different sites that manifest the CLSs are evaluated as $\phi_a = \phi_e =0$, $\phi_b=\phi_d=-1/(t_1+t_2) \: \phi_c$, when $E=\varepsilon_c$ and $t=1$ as illustrated in Fig.~\ref{five-site-model} (a). The missing amplitudes at sites $A$ and $E$ are caused by a destructive interference that yields trapping of the particles by strictly restricting the wavefunction to a particular region.

Our analytical approach reveals a fascinating consequence -- the appearance of a zero-energy flat band when $\varepsilon_c = 0$. This can be further understood using the rank-nullity theorem in linear algebra~\cite{lieb1989two}, which states that, if $T: V \rightarrow W$ is a linear map between two finite-dimensional vector spaces, then $\mathrm{dim}(\mathrm{im}(T))+\mathrm{dim}(\mathrm{ker}(T))=\mathrm{dim}(V)$, where ``$\mathrm{im}$" and ``$\mathrm{ker}$" denote the image and the kernel, respectively. In other words, any matrix $M$ of order $m \times n$ invariably satisfies the relation $\text{rank}(M) + \text{nullity}(M) = n$. Now, the system Hamiltonian $\mathcal{H}$ with two distinct sublattices $\alpha$ and $\beta$ can be expressed as

\begin{equation}
    \mathcal{H} = \begin{pmatrix}
0_{N_{\alpha} \times N_{\alpha}} & \mathcal{M}^{\dagger}_{N_{\alpha} \times N_{\beta}} \\
\mathcal{M}_{N_{\beta} \times N_{\alpha}} & 0_{N_{\beta} \times N_{\beta}}
\end{pmatrix}.
\end{equation}

Here, $N_{\alpha}$ and $N_{\beta}$ denote the number of $\alpha$-type and $\beta$-type sites in each unit cell, respectively. The five-site lattice [given in Eq.~\ref{eq:nhLieb}] consists of two triply-coordinated sites ($A$ and $E$) and three doubly-coordinated sites ($B$, $C$, and $D$). Hence, in our case, $N_{\alpha}=2$ and $N_{\beta} = 3$. Based on the above discussion, we can immediately write down the following set of equations relating rank, $R$, and nullity, $\Phi$,~\cite{sutherland1986localization}

\begin{gather}
   R(\mathcal{M}) + \Phi(\mathcal{M}) = N_{\alpha}=2, \nonumber \\
    R(\mathcal{H}) + \Phi(\mathcal{H}) = N_{\alpha}+N_{\beta}= 5, \nonumber \\
     R(\mathcal{H}) = R(\mathcal{M}) + R(\mathcal{M}^{\dagger}).     
\label{eq:rnknlty}
\end{gather} 

Using Eq.~\ref{eq:rnknlty} and the relation $R(\mathcal{M})=R(\mathcal{M}^{\dagger})$, it is straightforward to obtain $\Phi(\mathcal{H})= 2 \: \Phi(\mathcal{M})+1$. Additionally, the nullity of the non-singular matrix $\mathcal{M}$ is zero (i.e., $\Phi(\mathcal{M})=0$), which leads to $\Phi(\mathcal{H}) = 1$. This shows the emergence of a single zero-energy flat band, which remains pinned to sublattice $\beta$. In the present case, the above discussion leads to the amplitude distribution of the CLS, and the non-zero values are obtained only on sublattices ($B$, $C$, and $D$) [Fig.~\ref{five-site-model}(a)].

In contrast, the other solution $E_{-}$ given in Eq.~\ref{eq:solutions} offers a quadratic dispersion relation from which two low-energy bands can be obtained using the relation $|E^2-\varepsilon_{c}^{2}|/t^2<<1$. The expression for the dispersion relation of the two low-energy bands is $E=\pm[\varepsilon_{c}^{2}+t_{1}^{2}+t_{2}^{2}+2t_1t_2 \: \cos(k)]^{1/2}$. We note that the above expression resembles the band dispersion of a massive SSH chain with a complex mass term $\pm \varepsilon_c$. Moreover, through an appropriate choice of the momentum-dependent on-site potential ($\varepsilon_c=i \sqrt{\gamma^2 + 2 i \gamma \sin{k}}$), one can simply map it to a non-reciprocal SSH model that features rich non-equilibrium topological phases and interesting many body physics~\cite{banerjee2022chiral}. It is important to note that this decimation scheme is not a unique choice for mapping the five-site system to an identical two-level lattice. To establish this, we have alternatively eliminated all the doubly-coordinated sites $B$, $C$, and $D$, resulting in an equivalent two-level ladder system consisting of only the originally triply coordinated $A$ and $E$ sites. Similar to the previous case, the decimated ladder network can be decoupled into a massive SSH chain where alternating sites have the onsite energies $+\varepsilon_c$ and $-\varepsilon_c$, and an array of isolated lattice sites with locally pinned eigenfunctions (see supplement for a detailed analysis~\cite{supplement}). Therefore, the decimation scheme validates our hypothesis -- in the low energy limit, any quasi-one-dimensional bipartite non-Hermitian system exhibiting a flat band can be decoupled into a non-Hermitian SSH chain and periodically arranged isolated sites. The non-Hermitian SSH chain accounts for the band topology of the original lattice model, while the chain of isolated atoms manifests as flat band and CLS. We have further verified the above hypothesis for other non-Hermitian lattice systems with flat bands, namely, stub and diamond lattices (see supplement for detailed analysis~\cite{supplement}).

\section{Summary and outlook}

In conclusion, we have demonstrated the utility of the real space decimation scheme in elucidating various aspects of non-Hermitian spectral topology and emergent flat band physics. In particular, our systematic approach efficiently characterizes non-Hermitian phases and their transitions in complex multiband systems. Furthermore, we employ our formalism to study the skin mode physics, van Hove singularities using GBZ formulation, and transmittance properties in disordered/defect-induced non-Hermitian chains. Our approach has also shed light on CLSs in non-Hermitian systems, suggesting a hypothesis that quasi-one-dimensional non-Hermitian systems with flat bands can be decoupled into a non-Hermitian SSH chain and periodically arranged isolated sites. We note that, in a very recent work, the notion of biorthogonal polarization has been used to characterize the nontrivial topology of these systems~\cite{martinez2023topological}. 

While we have illustrated the decimation scheme for one-dimensional non-Hermitian problems, we envisage that it should be possible to generalize this to higher dimensional cases, including the study of dislocation-induced skin and anti-skin effects in two dimensions~\cite{bhargava2021non,schindler2021dislocation,panigrahi2022non}. Since our approach allows treating the disorder in an efficient way, we also expect our analytical approach to be useful in studying the scale invariance and flat band physics in non-Hermitian fractal networks~\cite{amir2015localization,manna2023inner}. It may also be interesting to employ our approach to use the renormalization flow equations to characterize critical phenomena and non-Hermitian phase transitions~\cite{zhou2020renormalization,rahul2022topological,lin2022topological,pi2021phase}. Our work, bridging the realms of real space renormalization schemes and non-Hermitian phenomena, is particularly timely given the surge of interest in non-Hermitian systems. We hope that our framework enables further work in this arena.

\section*{Acknowledgments}

A. Banerjee and R.S. are supported by the Prime Minister's Research Fellowship (PMRF). A. Bandyopadhyay acknowledges financial support IoE postdoctoral fellowship. A.N. acknowledges support from the Indian Institute of Science (startup grant SG/MHRD-19-0001).

 \bibliography{references}

\end{document}